\documentclass[conference]{IEEEtran}

\usepackage[T1]{fontenc}
\usepackage[utf8]{inputenc}
\usepackage{microtype}

\usepackage{libertinus}
\usepackage[libertine]{newtxmath}

\usepackage{amsmath}
\usepackage{newtxtext,newtxmath}

\usepackage{xcolor}

\usepackage{graphicx}
\usepackage{booktabs}
\usepackage{multirow}
\usepackage{array}
\usepackage{tabularx}
\usepackage{siunitx}

\usepackage{algorithm}
\usepackage{algpseudocode}

\usepackage{cite}
\usepackage{xcolor}
\usepackage{enumitem}
\usepackage{float}
\usepackage{placeins}
\usepackage{balance}

\usepackage{tikz}
\usetikzlibrary{positioning,arrows.meta,calc}

\usepackage{quantikz}

\usepackage[caption=false,font=footnotesize]{subfig}

\usepackage[
    colorlinks=true,
    linkcolor=blue,
    citecolor=blue,
    urlcolor=blue
]{hyperref}

\usepackage{tcolorbox}
\usepackage{booktabs}

\title{Distributed Quantum Error Correction with Bivariate Bicycle Codes in a Modular Architecture}

\author{
\IEEEauthorblockN{
Nitish Kumar Chandra\IEEEauthorrefmark{1}\IEEEauthorrefmark{2},
Eneet Kaur\IEEEauthorrefmark{2},
Reza Nejabati\IEEEauthorrefmark{2},
and Kaushik P. Seshadreesan\IEEEauthorrefmark{1}
}

\IEEEauthorblockA{\IEEEauthorrefmark{1}
Department of Informatics \& Networked Systems, School of Computing \& Information\\
University of Pittsburgh, Pittsburgh, PA 15260, USA\\
Emails: nkc16@pitt.edu, kausesh@pitt.edu
}

\IEEEauthorblockA{\IEEEauthorrefmark{2}
Quantum Labs, Cisco Systems, 3232 Nebraska Ave, Santa Monica, California 90404, USA\\
Emails: ekaur@cisco.com, rnejabat@cisco.com
}
}

\begin{document}
\maketitle

\begingroup \renewcommand\thefootnote{} \footnotetext{\textbf{This work has been submitted to the IEEE for possible publication. Copyright may be transferred without notice, after which this version may no longer be accessible.}} \endgroup

\begin{abstract}
Quantum low density parity check (qLDPC) codes, particularly bivariate bicycle (BB) codes, achieve competitive fault tolerance thresholds while offering substantially higher encoding rates than planar surface codes. However, their intrinsically long-range stabilizer structure makes them difficult to implement on monolithic devices with nearest neighbor connectivity and limited qubit capacity. In this work, we study the realization of a BB code in a modular multiprocessor architecture, where quantum processors are interconnected through shared Bell pairs. We consider processors with all to all internal connectivity, which is feasible on trapped ion and neutral atom platforms, enabling flexible local gate execution while inter-processor (non-local) gates are mediated by shared entanglement. We describe a star network architecture that can realize this distributed setting. We partition the qubits of the $[[144,12,12]]$ BB code across $4$, $6$, and $12$ quantum processors and analyze the resulting logical error rates and pseudo-threshold performance under circuit level noise by varying the number of processors and a scaling factor that captures the additional noise associated with nonlocal operations. We use Monte Carlo simulations with BP+OSD decoding and extend the previously known BB code ansatz to the distributed setting. Our results provide architectural insight and design considerations for distributed BB codes in modular quantum computing architectures.

\end{abstract}
\vspace{2pt}

\section{Introduction}

The realization of large scale fault-tolerant quantum computing will require \emph{modular architectures}, in which computation is distributed across smaller interconnected quantum processors rather than confined to a single chip or trap~\cite{Riel2026,yoder2025tourgrossmodularquantum,sakuma2025qflyopticalinterconnectmodular,Larasati2025,knorzer2025distributedquantuminformationprocessing}. In such architectures, quantum processing units are typically connected through \emph{photonic interconnects}, making the performance of the composite system depend not only on local gate fidelities but also on the quality of intermodule operations~\cite{Monroe2014,Ramette2024, NKChandra2025,Chandra2025,Chandra2026}. Although modularity provides a promising route to scalability, it also introduces additional noise through imperfect interconnects, making error suppression more challenging for reliable quantum computation. This motivates the use of \emph{quantum error correcting codes} to encode and protect logical information~\cite{deMacedoGuedes2026}. Consequently, \emph{distributed quantum error correction} has become an active area of recent study, since a code that performs well on an ideal monolithic device may behave quite differently when its stabilizer measurements are implemented across multiple quantum processors~\cite{babaie2024distributedquantumerrorcorrection,Chandra2025,clayton2025distributedquantumerrorcorrection,Sutcliffe2025}.


Within the broader family of \emph{quantum low density parity check} (qLDPC) codes, \emph{bivariate bicycle} (BB) codes have emerged as a promising class because they combine sparse stabilizers with nonzero encoding rate and competitive circuit level noise thresholds~\cite{Bravyi2024}. A notable example is the [[144,12,12]] BB code, which shows strong logical error suppression while achieving performance comparable to the planar surface code~\cite{Bravyi2024,yoder2025tourgrossmodularquantum}. Despite these advantages, the practical implementation of BB codes is constrained by hardware connectivity. In contrast to planar surface codes, whose syndrome extraction circuits naturally match two dimensional nearest-neighbor layouts, BB codes generally require geometrically nonlocal interactions. On current monolithic architectures, these interactions can increase routing overhead and circuit depth, which can degrade code performance~\cite{Berthusen2025,zhao2025simpleuniversalroutingstrategy}.

These architectural constraints make \emph{modular quantum architectures} a compelling setting for implementing \emph{BB codes}. This setting is relevant to trapped ion modules~\cite{Ye2025} connected by photonic links and to neutral atom quantum architectures~\cite{Wang2026}, where each processor can support flexible or programmable intramodule connectivity, while remote interactions are realized through entanglement assisted operations~\cite{Sinclair2025}. However, the reliance on remote entanglement introduces additional noise. As a result, performance becomes sensitive to both the number of partitions used to distribute the code across quantum processors and the noise associated with nonlocal operations. Increasing the number of partitions can introduce more inter-processor interactions, thereby adding noise to the system. This motivates the present work, which examines how partitioning and noisy nonlocal gates jointly affect BB code performance in a modular architecture.

\vspace{2pt}
Recent work has begun to examine how qLDPC codes can be adapted to practical hardware constraints by addressing issues of implementability tailored to specific platforms~\cite{vasic2025quantumlowdensityparitycheckcodes,mathews2025placingroutingquantumldpc,webster2026pinnaclearchitecturereducingcost,Wang2026,yang2026spacetimeefficienthardwarecompatiblecomplexquantum}.
 Poole \emph{et al.} proposed a neutral atom implementation of the $[[144,12,12]]$ BB code based on an optimized qubit layout together with long-range Rydberg gates \cite{Poole2025}. Berthusen \emph{et al.} proposed a bilayer architecture for realizing qLDPC codes using only two-dimensional local gates and classical communication, and showed that BB codes are compatible with its parallel syndrome-extraction scheme and routing strategy~\cite{Berthusen2025}. Shaw and Terhal introduced morphing circuits for BB codes that reduce the required qubit connectivity from degree six to five while maintaining competitive performance \cite{Shaw2025}.
\vspace{2pt}

Related efforts have also explored modular and distributed realizations of qLDPC codes. Strikis and Berent developed a framework for constructing qLDPC codes tailored to modular architectures~\cite{Strikis2023}. Yoder \emph{et al.} proposed a modular quantum computing architecture based on BB codes and discussed a compilation strategy adapted to its architectural constraints, with the goal of enabling universal fault-tolerant quantum computation~\cite{yoder2025tourgrossmodularquantum}. Tham \emph{et al.} studied distributed fault-tolerant quantum memories on a \(2 \times L\) array of qubit modules, including a sparse cyclic layout specialized to the \( [[144,12,12]] \) BB code~\cite{tham2025distributedfaulttolerantquantummemories}. Together, these works reflect growing interest in higher-rate qLDPC codes beyond the planar surface code and in their implementation on modular and distributed quantum architectures.
\vspace{2pt}

In this work, we investigate distributed realizations of the \( [[144,12,12]] \) BB code across multiple quantum processors. We study how code partitioning and nonlocal operation noise affect its logical error rates and pseudo-threshold behavior. The main contributions of this work are as follows:

\begin{itemize}
  \vspace{2pt}
    \item We consider a distributed implementation of the \( [[144,12,12]] \) BB code  across $4$, $6$, and $12$ quantum processing units (QPUs) in which QPUs use entanglement as a resource for remote gate operations, and  describe a star network architecture to realize this setting.
 \vspace{4pt}
    \item We define a circuit-level noise model that distinguishes \emph{local} and \emph{nonlocal} operations by assigning nonlocal gates an error rate scaled by \(\alpha \in \{1,3,5,7\}\) relative to local two qubit gates.
 \vspace{4pt}
    \item Using Monte Carlo simulations together with BP+OSD decoding, we investigate how the logical error rate and pseudo-threshold behavior vary with the number of QPUs and the strength of nonlocal noise.
 \vspace{4pt}
    \item We introduce a modified ansatz that extends the baseline BB-code ansatz to capture the additional noise in the distributed setting, enabling a systematic characterization of logical error rates across different nonlocal gate noise scalings and extrapolation to lower physical error rates.
\end{itemize}






In Sec.~\ref{bg}, we present the background theory relevant to this work. In Sec.~\ref{sec:architecture}, we describe the method used to partition the BB144 code across multiple QPUs and a star network architecture that can be used to realize this distributed setting. In Sec.~\ref{architecture modelling}, we discuss the noise model and simulation procedure used in our study. Sec.~\ref{results} contains the logical error-rate plots and pseudo-threshold estimates for different QPU partitions and nonlocal-gate noise scalings, along with the fitting ansatz used to characterize the threshold behavior of the BB code.  Finally, in Sec.~\ref{sec:conclusion}, we summarize the key findings and outline directions for future research.




   

\section{Background Theory}\label{bg}

This section reviews the \([[144,12,12]]\) bivariate bicycle (BB) code used in this work, including its construction, geometry, syndrome extraction, and decoding.

\subsection{Bivariate Bicycle Code Construction}

\emph{Bivariate bicycle} (BB) codes form a prominent family of CSS quantum LDPC codes that combine sparse stabilizer structure with finite encoding rate~\cite{Bravyi2024}. They can be defined on a periodic two-dimensional lattice of size \(\ell \times m\) and described by the quotient ring,
\begin{equation}
\mathbb{F}_2[x,y]/(x^\ell-1,\; y^m-1),
\end{equation}
where \(\mathbb{F}_2\) denotes the binary field and \(x,y\) represent cyclic shifts along the two lattice directions. Algebraically, a BB code is specified by two sparse binary circulant matrices, \(A\) and \(B\). In a commonly studied weight-6 subclass,
\begin{equation}
A(x,y)=x^{a_1}+y^{a_2}+y^{a_3},
\qquad
B(x,y)=y^{b_1}+x^{b_2}+x^{b_3},
\end{equation}
where \(a_i\) and \(b_i\) are integer shift parameters, arithmetic is performed over \(\mathbb{F}_2\), and the exponents of \(x\) and \(y\) are interpreted modulo \(\ell\) and \(m\), respectively.

\vspace{2pt}

Since BB codes are CSS codes, their stabilizers can be described by separate \(X\) and \(Z\)-type parity check matrices,
\begin{equation}
H_X = [\,A \mid B\,],
\qquad
H_Z = [\,B^\top \mid A^\top\,].
\label{eq:bb_pcm}
\end{equation}
The sparsity of \(A\) and \(B\) gives bounded weight checks, resulting in the low density parity check structure.

\subsection{Polynomial and connectivity description of the \texorpdfstring{$[[144,12,12]]$}{[[144,12,12]]} BB code}

In this work, we consider the \([[144,12,12]]\) bivariate bicycle (BB) code following the notation of  Ref.~\cite{Bravyi2024}. 
Its lattice dimensions are \((\ell,m)=(12,6)\), and the defining polynomials are,
\begin{equation}
A = x^3 + y + y^2,
\qquad
B = y^3 + x + x^2.
\label{eq:AB_144}
\end{equation}
We label the qubits and checks of the \([[144,12,12]]\) BB code by monomials on a \(12\times 6\) torus. Specifically,
\[
M=\{x^i y^j : i\in\mathbb{Z}_{12},\; j\in\mathbb{Z}_{6}\},
\]
so each pair \((i,j)\) identifies a lattice cell, with indices taken modulo \(12\) in the \(x\) direction and modulo \(6\) in the \(y\) direction. For each cell \((i,j)\), we denote the corresponding left and right data qubits by \(L_{i,j}\) and \(R_{i,j}\), and the associated \(X\) and \(Z\) checks by \(X_{i,j}\) and \(Z_{i,j}\).

\vspace{4pt}

The check connectivity follows directly from the polynomial form of the parity check matrices. From \(H_X=[A\mid B]\), an \(X\) check labeled by the monomial \(x^i y^j\) is connected to the left data qubits obtained by multiplying \(x^i y^j\) by each term of \(A\), and to the right data qubits obtained by multiplying \(x^i y^j\) by each term of \(B\). Using Eq.~\eqref{eq:AB_144}, 
\begin{align}
x^i y^j A
&= x^i y^j (x^3 + y + y^2) \nonumber \\
&= x^{i+3} y^j + x^i y^{j+1} + x^i y^{j+2},
\end{align}
and
\begin{align}
x^i y^j B
&= x^i y^j (y^3 + x + x^2) \nonumber \\
&= x^i y^{j+3} + x^{i+1} y^j + x^{i+2} y^j.
\end{align}
Hence,
\begin{equation}
X_{i,j}
\leftrightarrow
\{
L_{i+3,j},\,
L_{i,j+1},\,
L_{i,j+2},\,
R_{i,j+3},\,
R_{i+1,j},\,
R_{i+2,j}
\},
\label{eq:Xconnect_144}
\end{equation}
where all indices are taken modulo \((12,6)\). 

\vspace{4pt}


The \(Z\)-check connectivity is obtained similarly from \(H_Z=[B^T\mid A^T]\). Since transposition reverses the monomial exponents, 
\begin{equation}
B^T = y^{-3} + x^{-1} + x^{-2},
\qquad
A^T = x^{-3} + y^{-1} + y^{-2}.
\label{eq:ATBT_144}
\end{equation}
Therefore, a \(Z\) check labeled by \(x^i y^j\) is connected to the left data qubits obtained from \(x^i y^j B^T\) and to the right data qubits obtained from \(x^i y^j A^T\). Thus,
\begin{align}
x^i y^j B^T
&= x^i y^j (y^{-3} + x^{-1} + x^{-2}) \nonumber \\
&= x^i y^{j-3} + x^{i-1} y^j + x^{i-2} y^j,
\end{align}
and
\begin{align}
x^i y^j A^T
&= x^i y^j (x^{-3} + y^{-1} + y^{-2}) \nonumber \\
&= x^{i-3} y^j + x^i y^{j-1} + x^i y^{j-2}.
\end{align}
Accordingly,
\begin{equation}
Z_{i,j}
\leftrightarrow
\{
L_{i,j-3},\,
L_{i-1,j},\,
L_{i-2,j},\,
R_{i-3,j},\,
R_{i,j-1},\,
R_{i,j-2}
\},
\label{eq:Zconnect_144}
\end{equation}
 with indices taken modulo \((12,6)\). 
 
\vspace{4pt}

The algebraic construction above determines the connectivity uniquely, but it does not fix how the four objects \(L_{i,j}\), \(R_{i,j}\), \(X_{i,j}\), and \(Z_{i,j}\) are positioned within a graphical unit cell. For visualization, we adopt a fixed layout in which \(L\) is placed at the upper left, \(X\) at the upper right, \(Z\) at the lower left, and \(R\) at the lower right (See Fig.~\ref{fig:bb144_qpu6_overlay}). This choice affects only the graphical presentation; the connectivity of the code is still determined entirely by Eqs.~\eqref{eq:Xconnect_144} and~\eqref{eq:Zconnect_144}, consistent with the polynomial construction of Ref.~\cite{Bravyi2024}.

\subsection{Syndrome Extraction Schedule}\label{sec:schedule}

A full stabilizer measurement cycle is implemented in eight time steps~\cite{Bravyi2024}. The circuit contains seven rounds of CNOT gates, arranged so that each \(X\)-check and \(Z\)-check ancilla couples exactly once to each of its six neighboring data qubits.

\begin{figure}[h!]
    \centering
    \includegraphics[width=1.1\columnwidth]{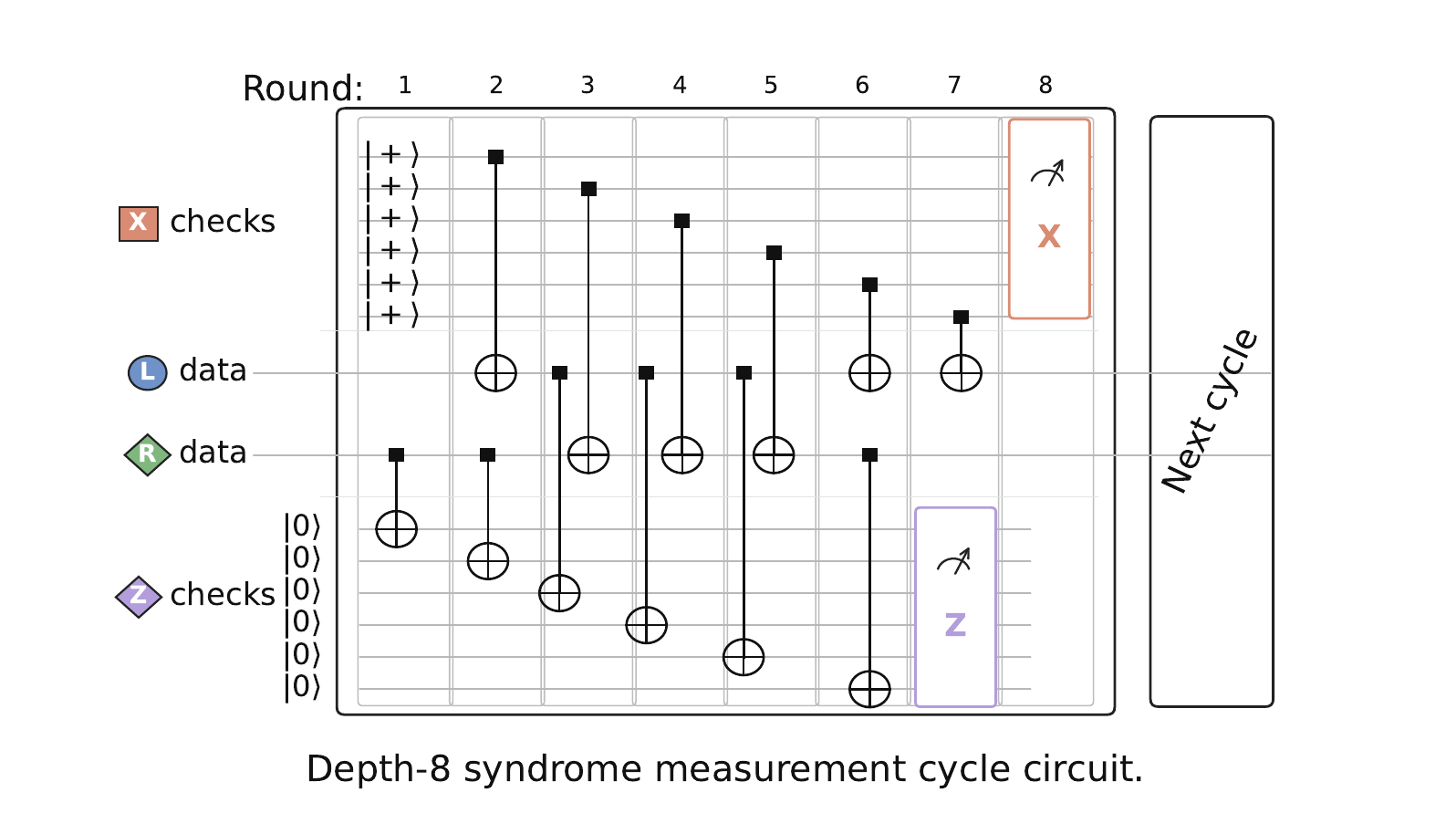}
    \caption{Depth-8 syndrome measurement cycle circuit.}
    \label{fig:depth8_cycle}
\end{figure}

\vspace{2pt}

The cycle proceeds as follows:

\begin{itemize}[leftmargin=*]

\item \textbf{Round 1:} All \(X\)-check ancillas are prepared in the \(|+\rangle\) state, and the first round of CNOT gates associated with the \(Z\)-check couplings is applied.

\item \textbf{Rounds 2--6:} In each round, one round of CNOT gates is applied from the \(X\)-check ancillas to the data qubits, and one round of CNOT gates is applied from the data qubits to the \(Z\)-check ancillas. 




\item \textbf{Round 7:} All \(Z\)-check ancillas are measured in the computational basis, and the final round of CNOT gates associated with the remaining \(X\)-check coupling is applied.


\item \textbf{Round 8:} All \(X\)-check ancillas are measured in the \(X\) basis, and all \(Z\)-check ancillas are prepared in the \(|0\rangle\) state for the next cycle. The data qubits are idle during this round.
\end{itemize}

\vspace{4pt}
The resulting syndrome extraction circuit therefore has constant depth, independent of the lattice size.

\subsection{Entanglement Assisted Nonlocal Gate}\label{nonlocalgate}

In distributed quantum computing, shared Bell pairs between two QPUs provide the entanglement resource needed to implement nonlocal gates. We can use communication qubits $a$ and $b$ in two QPUs to prepare the Bell state,
\begin{equation}
\ket{\Phi^+}_{ab}=\frac{1}{\sqrt{2}}(\ket{00}+\ket{11}).
\end{equation}

\begin{figure}[h!]
    \centering
    \includegraphics[width=0.95\columnwidth]{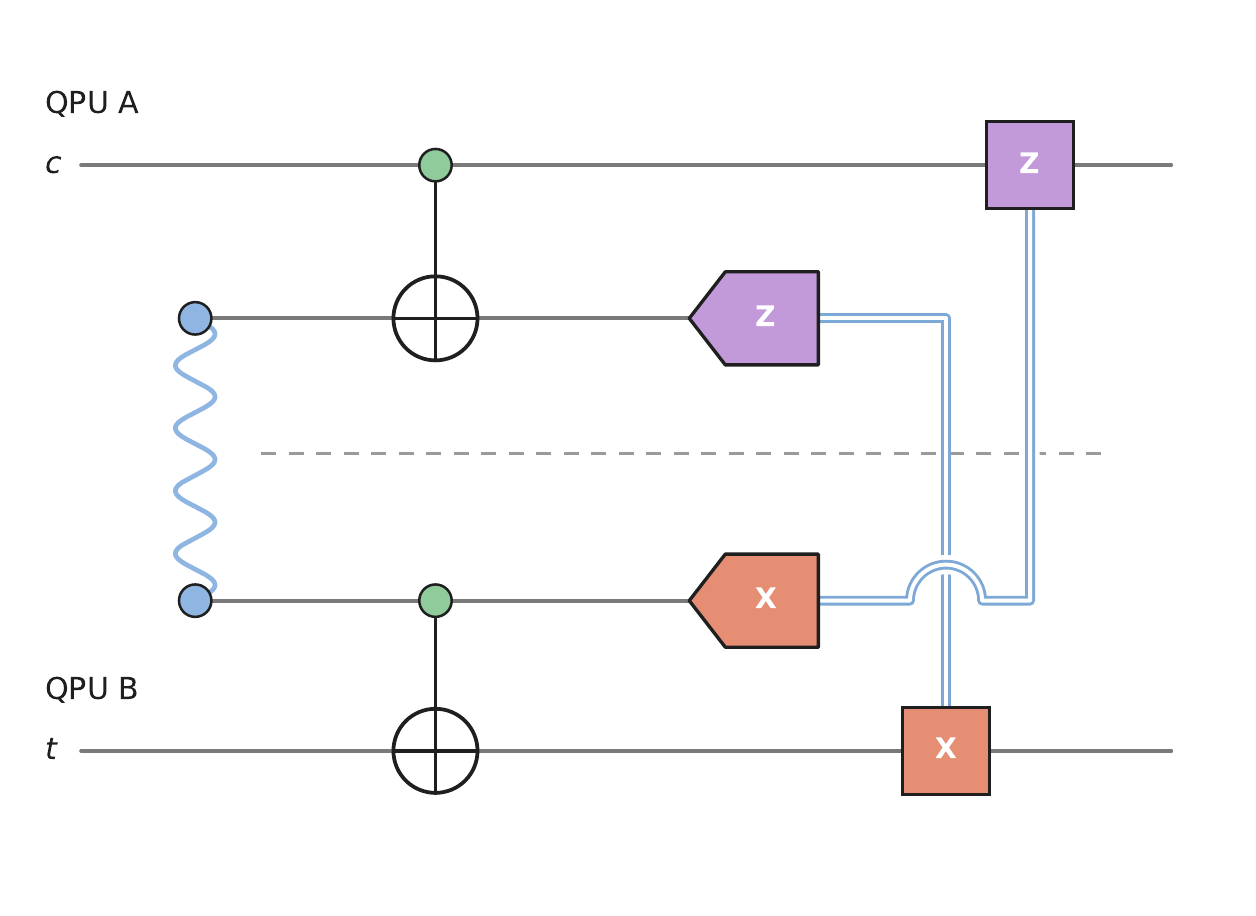}
    \caption{Remote CNOT between QPU A and QPU B using a shared Bell pair. Here, $c$ is the control qubit, $t$ is the target qubit, and $a,b$ are communication qubits that share the Bell pair.}
    \label{fig:remote_cnot_bell_pair}
\end{figure}

As illustrated in Fig.~\ref{fig:remote_cnot_bell_pair}, a CNOT gate between control qubit $c$ in QPU A and target qubit $t$ in QPU B is implemented using the shared Bell pair between communication qubits $a$ and $b$. We first apply a local CNOT from $c$ to $a$ and another local CNOT from $b$ to $t$. We then measure $a$ in the computational basis $\{\ket{0},\ket{1}\}$ and $b$ in the $X$ basis $\{\ket{+},\ket{-}\}$. The resulting classical outcomes specify the Pauli correction operations required to obtain the intended remote CNOT~\cite{Chou2018}. We use this approach to implement nonlocal gates during the syndrome measurement cycle shown in Fig.~\ref{fig:depth8_cycle}, whenever a CNOT acts on two qubits located in different QPUs.

\subsection{BP+OSD decoding}
\label{subsec:bposd_decoding}

Belief propagation with ordered statistics decoding (BP+OSD) is a decoding method for sparse parity-check codes~\cite{Roffe_LDPC_Python_tools_2022,roffe_decoding_2020}. Given a measured syndrome, belief propagation estimates the most likely error events by iteratively passing probabilistic messages along the Tanner graph of the code. For quantum CSS codes, the two error sectors can be decoded separately: \(Z\)-type errors are inferred from \(X\)-type syndrome information, while \(X\)-type errors are inferred from \(Z\)-type syndrome information.

When the Tanner graph contains many short cycles, messages passed along different paths can become correlated, reducing the accuracy of the BP estimate. Ordered statistics decoding is therefore used as a postprocessing step: it uses the soft reliability information produced by belief propagation to search for a correction that is consistent with the measured syndrome. In this work, we decode the BB code using BP+OSD, following the circuit-level simulation workflow of Refs.~\cite{Bravyi2024,BravyiBivariateBicycleCodesGitHub}.

\section{Modular Distributed Architecture for Bivariate Bicycle Codes}
\label{sec:architecture}

We partition the \([[144,12,12]]\) BB code across multiple QPUs and describe a star network architecture to realize the resulting distributed layout.

\begin{figure*}
[h!]
    \centering
    \includegraphics[width=1.45\columnwidth]{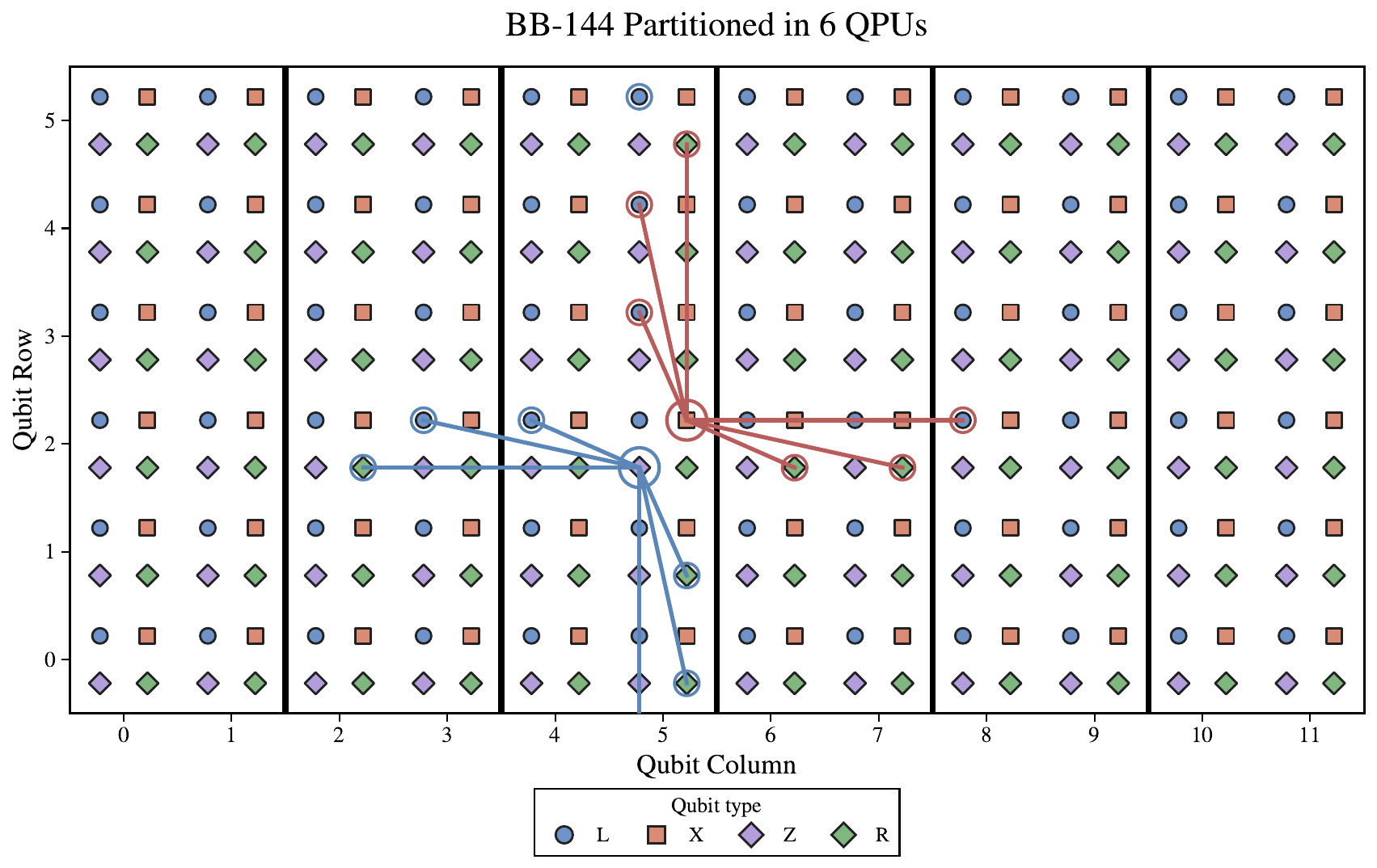}
    \caption{QPU partitioning of the \([[144,12,12]]\) bivariate bicycle code for \(N_{\mathrm{QPU}}=6\). The \(12\) \(x\)-columns are divided into six blocks of width \(c=2\), and the highlighted cell \((5,2)\) lies in QPU 2. The markers show the qubit types \(L\), \(X\), \(Z\), and \(R\). The overlaid blue and red edges show the \(X\)- and \(Z\)-check connectivity of the highlighted cell, respectively, as defined in Eqs.~\eqref{eq:Xconnect_144} and~\eqref{eq:Zconnect_144}. Edges within QPU 2 are local, while edges crossing into other QPU blocks are nonlocal. The horizontal and vertical axes label the torus cell coordinates
\((x,y)\in\mathbb{Z}_{12}\times\mathbb{Z}_{6}\); each cell contains the
four qubits \(L\), \(X\), \(Z\), and \(R\).}
    \label{fig:bb144_qpu6_overlay}
    
\end{figure*}



\subsection{QPU Partitioning of the BB144 Code}

\vspace{4pt}
The $[\![144,12,12]\!]$ bivariate bicycle code (BB144) is defined on the discrete torus $\mathbb{Z}_\ell \times \mathbb{Z}_m$, with $\ell = 12$ and $m = 6$, so that the torus contains $\ell m = 72$ sites. Each site (cell) contains the set of qubits $\{L,R,X,Z\}$, corresponding respectively to the left data qubit, right data qubit, $X$ check qubit, and $Z$ check qubit. 



To index the site, we assign to each \textit{site} an integer label $i \in \{0,1,\ldots,71\}$ and map it to a two dimensional coordinate on the torus via,
\begin{equation}
\begin{aligned}
i &\longmapsto (x_i, y_i),\\
(x_i, y_i)
&=
\left(
\left\lfloor \frac{i}{m} \right\rfloor,\;
i \bmod m
\right).
\end{aligned}
\label{eq:torus_map}
\end{equation}
where $\lfloor i/m \rfloor$ is the floor of $i/m$ (the greatest integer less than or equal to $i/m$), and $i \bmod m$ is the remainder upon division of $i$ by $m$. The first quantity determines the $x$ coordinate and the second determines the $y$ coordinate. In this way, the integer label $i$ enumerates the \textit{sites}, while $(x_i,y_i)$ identifies the corresponding location on the $12 \times 6$ torus. We refer to $x_i$ as the \emph{x-column} of site $i$.

We now outline the partitioning rule, specify the locality classification, and consider the limiting cases.

\paragraph{Partition rule}
We partition the $\ell = 12$ columns into $N_\mathrm{QPU}$ contiguous blocks of equal width,
\begin{equation}
c = \frac{\ell}{N_\mathrm{QPU}},
\label{eq:block_width}
\end{equation}
so that each block is assigned to one QPU. For this partitioning to yield equal-sized blocks, $N_\mathrm{QPU}$ must divide $\ell$, and hence the valid choices for BB144 are,
\begin{equation}
N_\mathrm{QPU} \in \{1,2,3,4,6,12\}.
\end{equation}
The QPU label associated with site $i$ is then given by,

\begin{equation}
\begin{gathered}
\nu(i)= \left\lfloor \frac{x_i}{c} \right\rfloor \\[2pt]
= \left\lfloor
\frac{\lfloor i/m \rfloor}{\ell / N_\mathrm{QPU}}
\right\rfloor,
\qquad
\nu(i) \in \{0,1,\ldots,N_\mathrm{QPU}-1\}
\end{gathered}
\label{eq:qpu_assignment}
\end{equation}
Here, the same label \(\nu(i)\) is assigned to all four qubit types associated with site \(i\), namely \(\{L,R,X,Z\}\). Since \(\nu(i)\) depends only on the column coordinate \(x_i\), each QPU corresponds to a contiguous vertical strip of the torus containing \(c \times m\) sites.
For example if  $N_{\mathrm{QPU}}=6$, the partition width is,
\begin{equation}  c=\frac{\ell}{N_{\mathrm{QPU}}}=\frac{12}{6}=2,
\end{equation}
so each QPU contains two adjacent x-columns. The resulting partition is,
\begin{equation}
\begin{aligned}
    &\text{QPU 0}: \{0,1\}, \qquad \text{QPU 1}: \{2,3\},\\
    &\text{QPU 2}: \{4,5\}, \qquad \text{QPU 3}: \{6,7\},\\
    &\text{QPU 4}: \{8,9\}, \qquad \text{QPU 5}: \{10,11\}.
\end{aligned}
\end{equation}

For the highlighted site (cell) $(5,2)$ in Figure~\ref{fig:bb144_qpu6_overlay}, the x-coordinate is $x=5$, and therefore its QPU label is,

\begin{equation}
    \nu(5,2)=\left\lfloor \frac{5}{2} \right\rfloor=2.
\end{equation}
Hence, the selected cell belongs to QPU 2. Here we use the same assignment rule in coordinate form, \(\nu(x,y)=\left\lfloor x/c\right\rfloor\).

Figure~\ref{fig:bb144_qpu6_overlay} shows the \(N_{\mathrm{QPU}}=6\) partition described above, together with the connectivity associated with the highlighted cell. Connections whose endpoints lie within QPU 2 remain local, while connections that cross into other QPU blocks are treated as nonlocal.

\paragraph{Locality classification}
A two-qubit gate between cells $i$ and $j$ is classified according to
whether both endpoints (cells) reside on the same QPU:
\begin{equation}
\text{gate type}
=
\begin{cases}
\textit{local}, & \text{if } \nu(i)=\nu(j),\\
\textit{nonlocal}, & \text{if } \nu(i)\neq\nu(j).
\end{cases}
\label{eq:gate_type}
\end{equation}

Local gates are executed entirely within one quantum processing unit (QPU), whereas nonlocal gates require inter-QPU communication and are assigned a higher error rate. This classification forms the basis of the two-rate noise model described in Sec.~\ref{noise modelling}.


\paragraph{Limiting cases}
When \(N_\mathrm{QPU}=1\), all qubits are placed on a single QPU, so every CNOT gate is local and the model reduces to the standard circuit-level noise model used as the baseline. At the opposite extreme, \(N_\mathrm{QPU}=12\) gives the finest partition as per our partitioning method of the \([[144,12,12]]\) BB code, with one \(x\)-column assigned to each QPU \((c=1)\). This corresponds to 24 qubits per QPU and maximizes the number of nonlocal interactions.

\vspace{2pt}

We used a column-based partitioning because it follows the translational structure of the BB code and aims to minimize nonlocal interactions between QPUs. Further optimization could relax the cell-preserving constraint, for example, by allowing qubits within a cell to be split across different QPUs.

\subsection{A Modular Architecture}

\begin{figure}[h!]
    \centering
    \includegraphics[width=0.99\linewidth]{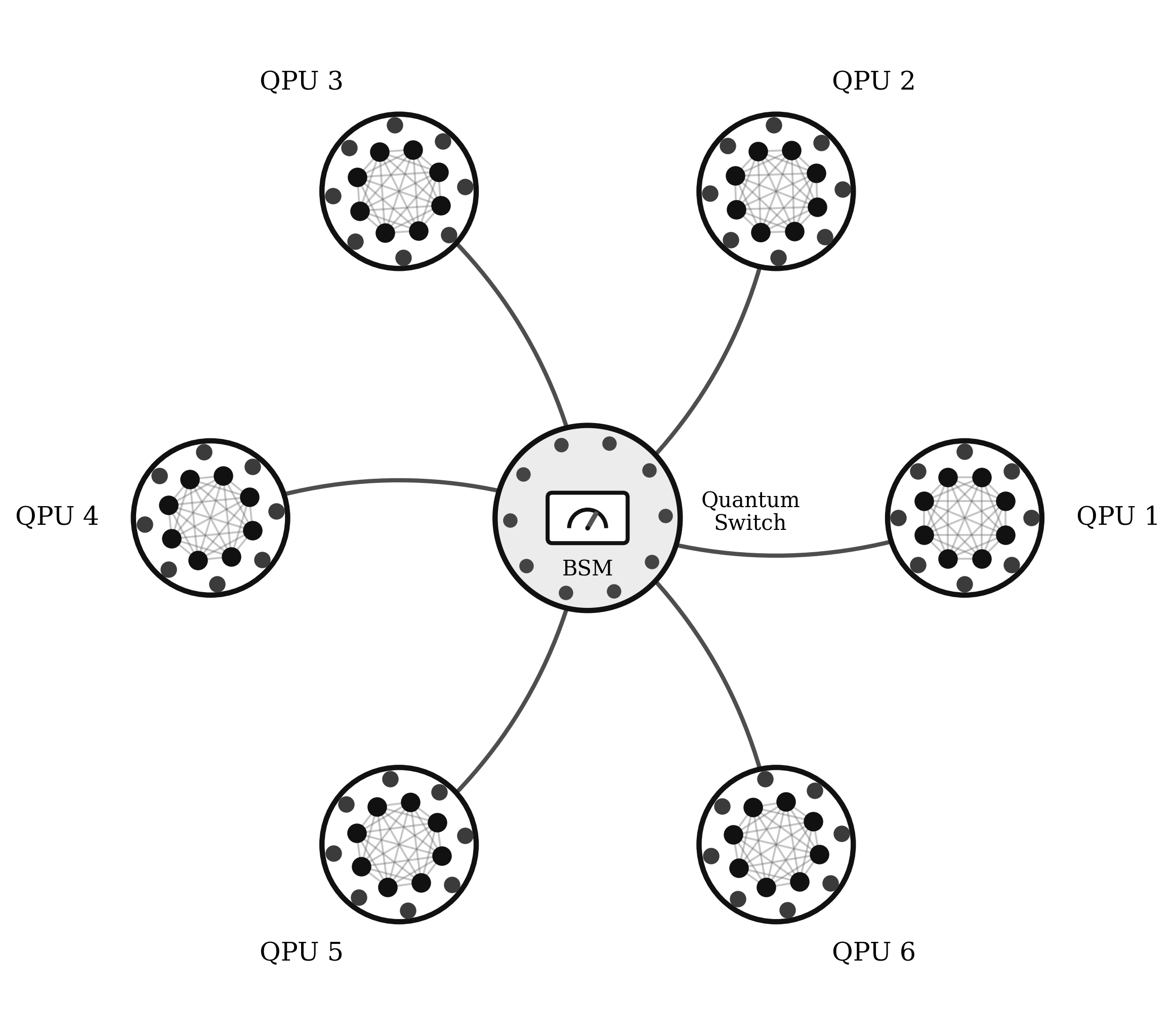}
    \caption{
    Star network architecture in which multiple QPUs (6) are connected through a central quantum switch. Each QPU contains data and ancilla qubits with intra-module connectivity, together with communication qubits used to generate entangled pairs with the switch. The switch uses these entangled pairs, together with Bell-state measurements (BSM), to enable nonlocal operations between QPUs.
    }
    \label{fig:distributed_architecture}
\end{figure}

We consider a modular quantum computing architecture composed of multiple Quantum Processing Units (QPUs) connected through a central quantum switch, as illustrated in Figure~\ref{fig:distributed_architecture}, forming a star network~\cite{Iesta2024}. This architecture provides a viable setting for implementing quantum low density parity check (QLDPC) codes, in particular bivariate bicycle (BB) codes, whose stabilizer measurements generally involve nonlocal interactions.

We assume that each QPU contains a fixed number of physical qubits and supports all-to-all connectivity among its qubits. The qubits within each QPU are divided into three categories: \textit{data qubits}, which store the encoded quantum information; \textit{ancilla qubits}, which are used for stabilizer measurements; and \textit{communication qubits}, which connect the QPU to the central switch. This allows data and ancilla qubits to interact locally with communication qubits and thereby participate, when required, in entanglement assisted nonlocal operations.

The central quantum switch serves as an entanglement generation and routing node. A quantum scheduler determines the timing of Bell pair generation and Bell state measurements, enabling the switch to establish Bell pairs with communication qubits in different QPUs and perform entanglement swapping to create shared entanglement between them~\cite{NChandra2024}. These entangled links are then used to realize the nonlocal entangling operations required for BB code syndrome extraction. Although we focus on this switch based star architecture as a specific realization, the distributed layout can be implemented in any networked architecture capable of generating Bell pairs between arbitrary QPUs. 

We also note that, while we have described all-to-all connectivity within each small QPU, BB syndrome extraction only requires specific check qubit and data qubit interactions. Thus, QPUs that provide these required connections are sufficient for implementing this code, even for the distributed realization.

\section{Architectural Noise Modelling and Simulations}\label{architecture modelling}

For the modular architecture shown in Figure~\ref{fig:distributed_architecture}, we model an elementary entanglement link between a QPU communication qubit and a switch qubit by a noisy two-qubit state \(\rho_{\mathrm{QS}}\) with fidelity,
\begin{equation}
F_{\mathrm{QS}}=\langle \Phi^{+}|\rho_{\mathrm{QS}}|\Phi^{+}\rangle < 1,
\end{equation}
where \(\ket{\Phi^+}\) is the ideal Bell state. The corresponding entanglement-link infidelity is \(\varepsilon_{\mathrm{QS}}=1-F_{\mathrm{QS}}\). 
Once the elementary entanglement links between the QPUs and the switch are generated, they can be used to create an end-to-end entanglement link between two remote QPUs.  
To achieve this, the switch performs entanglement swapping through a Bell-state measurement on its two qubits. If the elementary entanglement pairs and the swapping operation are imperfect, the resulting inter-QPU Bell pair has fidelity \(F_{\mathrm{swap}}<1\)~\cite{Briegel1998,Halder2024}.

\vspace{4pt}

The Bell pair obtained after entanglement swapping can be used to implement a nonlocal gate between qubits in different QPUs. We model this gate as an effective noisy operation with error rate,
\begin{equation}
\varepsilon_{\mathrm{NL}}
=
f\!\left(
\varepsilon_{\mathrm{swap}},
\varepsilon_{\mathrm{loc}},
\varepsilon_{\mathrm{meas}},
\varepsilon_{\mathrm{ff}},
\ldots
\right),
\end{equation}
where \(\varepsilon_{\mathrm{swap}}\) captures the error in the entanglement swapping operation. The remaining terms represent error contributions from local operations (loc), measurements (meas), feedforward and conditional corrections (ff), and other noise sources. The map \(f\) absorbs these contributions and depends on the implementation details.

\vspace{2pt}

An explicit expression for the nonlocal gate error is left to future work, as it requires a detailed noise analysis beyond the scope of this study. Since an effective error rate for the non-local gate can be derived once the noise model and implementation details are specified, we capture the combined impact of Bell-pair infidelity and other operational imperfections through a single effective noise parameter. Specifically, if \(p\) denotes the baseline error rate of a two-qubit gate within a QPU, then a two qubit nonlocal gate is assigned the elevated rate,
\begin{equation}
p_{\mathrm{NL}}=\alpha p,
\qquad \alpha \geq 1,
\end{equation}
where \(\alpha\) represents the cumulative noise penalty associated with nonlocal gate execution.

\subsection{Distributed circuit-level noise model}\label{noise modelling}

To model syndrome extraction in the distributed setting, we first assign each qubit to a QPU according to the partition map introduced above, and denote the QPU label of qubit \(q\) by \(\nu(q)\). Every CNOT gate in the syndrome extraction circuit is then classified according to the locations of its endpoint qubits.


We consider a circuit-level stochastic Pauli noise model with baseline physical error rate \(p\). Preparation, measurement, and idle errors are all assigned this same base rate. Our distributed realization modifies only the CNOT operations: for a CNOT acting on qubits \(q_1\) and \(q_2\), the total error probability is
\begin{equation}
r(q_1,q_2)=
\begin{cases}
p, & \nu(q_1)=\nu(q_2),\\[4pt]
\alpha p, & \nu(q_1)\neq \nu(q_2),
\end{cases}
\label{eq:cnot_rate}
\end{equation}
where \(\alpha \ge 1\) is a nonlocal noise penalty factor. Thus, the distributed model preserves the syndrome extraction schedule of the baseline circuit-level model, while assigning an elevated error rate to CNOT gates that cross QPU boundaries.

We consider the following noise souces:
\begin{itemize}[leftmargin=1.2em,labelsep=0.8em]
    \item 

An \textit{idle operation} is followed by a one qubit depolarizing error with probability \(p\),
\begin{equation}
\mathcal{E}_{\mathrm{idle}}(\rho)
=
(1-p)\rho
+
\frac{p}{3}
\left(
X\rho X + Y\rho Y + Z\rho Z
\right).
\label{eq:idle_channel}
\end{equation}
\item \textit{Preparation} and \textit{measurement errors} are modeled as basis dependent Pauli flips at rate \(p\): a \(\ket{+}\) preparation or an \(X\) basis measurement is followed by a \(Z\) error with probability \(p\), whereas a \(\ket{0}\) preparation or a \(Z\) basis measurement is followed by an \(X\) error with probability \(p\).

\vspace{2pt}
\item For a \textit{CNOT gate}, the error model is a two qubit depolarizing channel applied after the ideal gate. Let
\[
\mathcal{P}_2=\{I,X,Y,Z\}^{\otimes 2},
\qquad
\mathcal{P}_2^\ast=\mathcal{P}_2 \setminus \{I \otimes I\},
\]
where \(I \otimes I\) denotes the two qubit identity operator. If \(r=r(q_1,q_2)\) is the total CNOT error probability from Eq.~\eqref{eq:cnot_rate}, then
\begin{equation}
\begin{aligned}
\mathcal{E}_{\mathrm{CNOT}}^{(q_1,q_2)}(\rho)
&= (1-r)\rho \\
&\quad + \frac{r}{15}
\sum_{P \in \mathcal{P}_2^\ast} P \rho P^\dagger .
\end{aligned}
\label{eq:cnot_channel}
\end{equation}

Equivalently, conditioned on a CNOT error, one of the 15 nonidentity two qubit Pauli errors is sampled uniformly at random. Local and nonlocal CNOTs therefore differ only in the total error strength \(r\); the Pauli error classification is unchanged, but its overall weight is larger by a factor of \(\alpha\) for nonlocal gates.

\end{itemize}

For noise modelling within the CSS decomposition, \(X\) and \(Z\) components are treated separately.
For the one qubit depolarizing channel of Eq.~\eqref{eq:idle_channel}, the marginal probability of an \(X\) type error or a \(Z\) type error is,
\begin{equation}
p_{\mathrm{proj}}^{(1)}=\frac{2p}{3},
\label{eq:single_proj}
\end{equation}
because two of the three nonidentity single qubit Pauli errors contribute to each component: \(X\) and \(Y\) contribute to the \(X\) component, while \(Z\) and \(Y\) contribute to the \(Z\) component.

\vspace{2pt}

For a CNOT error with total rate \(r\), the two qubit depolarizing channel contains 15 nonidentity Pauli errors, each occurring with probability \(r/15\). After projection onto either the \(X\) component or the \(Z\) component, these 15 errors are grouped into three classes according to whether the induced component acts on the first qubit only, on the second qubit only, or on both qubits. For example, in the \(Z\) component, the first class is generated by \(\{ZI,YI,ZX,YX\}\), the second by \(\{IZ,IY,XZ,XY\}\), and the third by \(\{ZZ,YY,YZ,ZY\}\). Each class therefore contains four of the 15 nonidentity two qubit Pauli errors, so each has marginal probability,
\begin{equation}
p_{\mathrm{proj}}^{(2)}(r)=\frac{4r}{15}.
\label{eq:two_proj}
\end{equation}
The same reasoning applies to the \(X\) component. Substituting Eq.~\eqref{eq:cnot_rate} gives,
\begin{equation}
p_{\mathrm{proj,L}}^{(2)}=\frac{4p}{15},
\qquad
p_{\mathrm{proj,NL}}^{(2)}=\frac{4\alpha p}{15},
\label{eq:local_nonlocal_proj}
\end{equation}
for the local and nonlocal error contributions, respectively. This projection simplifies decoding by reducing each two-qubit depolarizing CNOT error to the effective \(X\) or \(Z\)-type error classes~\cite{BravyiBivariateBicycleCodesGitHub}.

\vspace{4pt}

\subsection{Distributed simulation workflow}
\label{sec:distributed_workflow}


Our simulations follow the workflow used by Bravyi \emph{et al.}, consisting of an \textit{offline} preprocessing stage and an\textit{ online} Monte Carlo stage, where the \(X\) and \(Z\)-error components are decoded separately using BP+OSD~\cite{Bravyi2024}. In the offline stage, the repeated BB-144 syndrome extraction circuit is constructed for the targeted number of measurement cycles, and effective error mechanisms are generated by propagating representative single-error events through the circuit. These events include the projected CSS-sector error classes associated with circuit-level noise and are used to construct the decoding matrices and effective channel probabilities for the separate \(X\) and \(Z\)-sector decoding problems.

During the online Monte Carlo stage, noisy syndrome-extraction circuits are sampled from the prescribed circuit-level noise model, and the resulting syndrome histories are decoded using the precomputed decoding matrices and channel probabilities. A trial is deemed successful if and only if both error sectors are decoded without a logical error; an incorrect inference in either sector contributes to the overall logical failure rate. 



The \textit{key extension} introduced in this work is the incorporation of a distributed architecture. After assigning qubits to QPUs using the partition map defined above, we classify each CNOT in the BB-144 syndrome-extraction circuit as local or nonlocal according to the locations of its endpoint qubits. In particular, the offline stage constructs the effective error mechanisms for the chosen values of \(p\), \(\alpha\), and \(N_{\mathrm{QPU}}\), assigning rate \(p\) to local CNOTs and \(\alpha p\) to nonlocal CNOTs. The online stage then samples noisy syndrome-extraction circuits using the same local and nonlocal error-rate assignments and decodes the resulting syndrome histories to determine whether each trial produces a logical error.

\section{Results}\label{results}

\label{sec:simulation_objectives}

Our objective is to quantify how QPU partitioning and elevated nonlocal noise affect the logical performance of the \([\![144,12,12]\!]\) BB code. We consider partitions across \(4\), \(6\), and \(12\) QPUs and use the parameter \(\alpha\) to scale the nonlocal operation error rate relative to the baseline circuit-level noise, with \(\alpha=1\) recovering the baseline case. For each \(\alpha \in \{1,3,5,7\}\), we simulate physical error rates in the range \(10^{-3}\) to \(10^{-2}\), using \(30{,}000\) Monte Carlo trials per data point.

\vspace{2pt}

For each QPU partition and each pair \((p,\alpha)\), we simulate the memory experiment for \(12\) syndrome-extraction cycles, estimate the total logical failure probability, and convert it to a logical error rate per cycle. This yields logical error curves \(p_L(p,\alpha)\) for each partition, which we use to compare the combined effects of code partitioning and nonlocal noise and to extract a pseudo-threshold for each setting.

\subsection{Proposed $\alpha$-dependent BB ansatz for logical error fitting}


To study how nonlocal gate noise affects BB-code performance, we use a one-parameter family of noise models indexed by \(\alpha\). Our goal is to extend the BB ansatz while preserving the structure of the original threshold analysis and allowing the logical error curve to vary systematically with \(\alpha\).

\vspace{4pt}

Let \(P_L(N_c)\) be the probability of observing a logical failure after \(N_c\) syndrome cycles. We convert this quantity to an effective per-cycle logical error rate by defining,
\begin{equation}
p_L
=
1-\bigl(1-P_L(N_c)\bigr)^{1/N_c}.
\label{eq:per_cycle_definition}
\end{equation}
Equivalently, \(p_L\) is the cycle-level failure probability that would reproduce the same total failure probability \(P_L(N_c)\) after \(N_c\) independent cycles.

For the baseline BB memory experiment, Bravyi \emph{et al.} model the per-cycle logical error rate using the expression~\cite{Bravyi2024},
\begin{equation}
p_L(p)
=
p^{d_{\mathrm{circ}}'/2}
\exp\!\left(c_0+c_1 p+c_2 p^2\right),
\label{eq:bb_baseline_ansatz}
\end{equation}
where \(p\) denotes the physical circuit-level error rate and \(c_0,c_1,c_2\) are fit parameters. For the \([[144,12,12]]\) BB code considered here, \(d_{\mathrm{circ}}'=10\) corresponds to the circuit-level distance upper bound reported in Ref.~\cite{Bravyi2024}. Here, circuit-level distance refers to the minimum number of faulty circuit locations required to produce an undetected logical failure during syndrome extraction.


Another key quantity used in this work is the \textit{pseudo-threshold}. It is defined as the break even physical error rate \(p_0\) satisfying,
\begin{equation}
p_L(p_0)=k p_0,  
\end{equation}
The term \(k p_0\) estimates the probability that at least one of the \(k\) unencoded qubits experiences an error.





\vspace{2pt}

When the nonlocal noise scale \(\alpha\) is introduced, our working \textit{assumption} is that \(\alpha\) changes the effective error rates of the nonlocal circuit locations, while leaving the underlying BB code and syndrome extraction circuit unchanged. We therefore keep the leading power \(p^{d_{\mathrm{circ}}'/2}\) fixed and allow only the fitted correction coefficients to depend on \(\alpha\).  Therefore, the generalized form of ansatz becomes,
\begin{equation}
p_L(p,\alpha)
=
p^{d_{\mathrm{circ}}'/2}
\exp\!\left(
c_0(\alpha)+c_1(\alpha)\,p+c_2(\alpha)\,p^2
\right).
\label{eq:alpha_generalized_ansatz}
\end{equation}


To anchor the model at the baseline case, we write,
\begin{equation}
c_j(1)=c_j^{(1)},
\qquad j\in\{0,1,2\},
\end{equation}
and expand each coefficient about $\alpha=1$. The most general local expansion is,
\begin{equation}
c_j(\alpha)
=
c_j^{(1)}
+
u_j(\alpha-1)
+
v_j(\alpha-1)^2
+\cdots.
\label{eq:alpha_taylor_expansion}
\end{equation}
A linear truncation retains only the first response of the fit coefficients to the nonlocal noise scale. However, this may be too restrictive. 
A quadratic truncation,
\begin{equation}
c_j(\alpha)
=
c_j^{(1)}
+
u_j(\alpha-1)
+
v_j(\alpha-1)^2,
\label{eq:quadratic_alpha_coefficients}
\end{equation}
is therefore a reasonable next step. It is flexible enough to capture nonlinear dependence on $\alpha$, yet still sufficiently structured to remain interpretable and closely tied to the original BB form.

\vspace{2pt}

Substituting Eq.~\eqref{eq:quadratic_alpha_coefficients} into Eq.~\eqref{eq:alpha_generalized_ansatz} yields the quadratic $\alpha$ dependent BB model,
\begin{equation}
\begin{split}
p_L(p,\alpha)
&=
p^{d_{\mathrm{circ}}'/2}
\exp\!\Big[
c_0^{(1)}+u_0(\alpha-1)+v_0(\alpha-1)^2 \\
&\quad
+\bigl(c_1^{(1)}+u_1(\alpha-1)+v_1(\alpha-1)^2\bigr)p \\
&\quad
+\bigl(c_2^{(1)}+u_2(\alpha-1)+v_2(\alpha-1)^2\bigr)p^2
\Big].
\end{split}
\label{eq:quadratic_alpha_bb_ansatz}
\end{equation}

This construction preserves the BB power law exponent, matches the baseline model exactly at $\alpha=1$, and allows the fitted curve to deform smoothly as the nonlocal noise scale is varied. The corresponding $\alpha$ dependent pseudo-threshold is defined similarly by,
\begin{equation}
p_L\bigl(p_0(\alpha),\alpha\bigr)=k\,p_0(\alpha).
\label{eq:alpha_dependent_pseudothreshold}
\end{equation}

The rationale behind this ansatz is as follows: the leading exponent is inherited from the BB circuit analysis and is kept fixed because \(\alpha\) changes only the relative strength of nonlocal errors, not the syndrome-extraction procedure. The subleading coefficients are allowed to depend on \(\alpha\), since they capture how the logical error curve changes as the nonlocal error rate is varied. Thus, the quadratic \(\alpha\)-dependent ansatz provides a \textit{structured extension} of the original BB ansatz.



\subsection{Pseudo-threshold fit results for 12 QPUs}

\begin{figure}[htbp]
    \centering
    \includegraphics[width=0.50\textwidth]{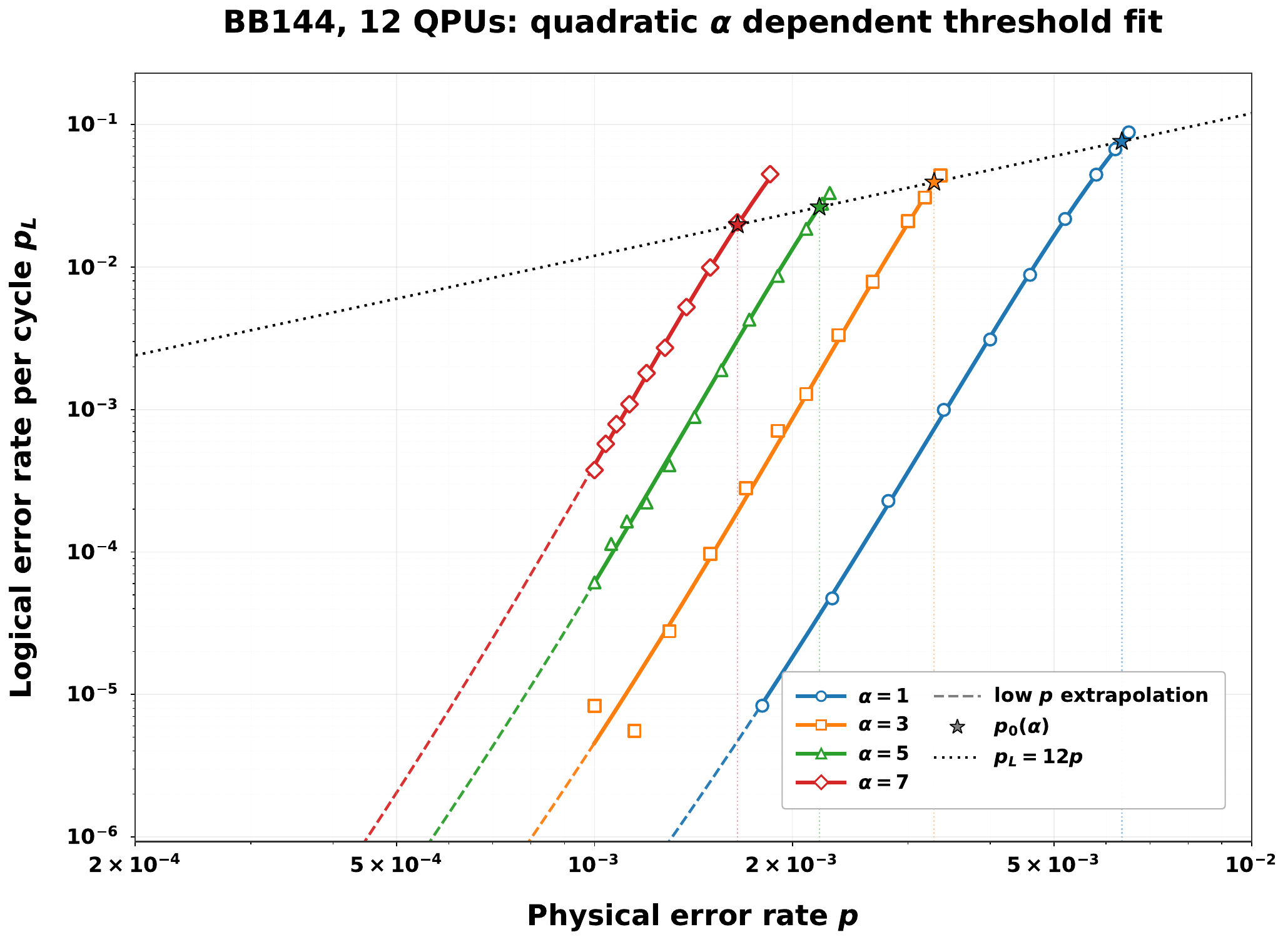}
    \caption{BB144 threshold fitting with quadratic alpha dependence. Logical error rate per cycle ($p_L$) vs. physical error rate ($p$) for different $\alpha$ values. The curves represent the BB ansatz fit, and asterisks mark the pseudo-thresholds $p_0(\alpha)$. The dashed lines show extrapolation to lower $p$.}
    \label{fig:bb144_threshold_fit}
\end{figure}


Figure~\ref{fig:bb144_threshold_fit} shows the quadratic \(\alpha\)-dependent fit of the BB logical-error ansatz for the BB-144 code distributed across 12 QPUs. The plot reports the logical error rate per cycle, \(p_L\), plotted against the physical error rate, \(p\), for different values of the nonlocal noise scale \(\alpha\). For each \(\alpha\), the solid colored curve shows the fitted ansatz over the data range used for fitting, while the dashed segment shows the extrapolation to lower \(p\). The dotted black line gives the break-even condition \(p_L=12p\), since the BB-144 code encodes \(k=12\) logical qubits. The marked intersections define the \(\alpha\)-dependent pseudo-thresholds \(p_0(\alpha)\) through Eq.~\eqref{eq:alpha_dependent_pseudothreshold}.
\vspace{2pt}
As \(\alpha\) increases, the fitted logical-error curves shift upward, indicating a higher logical error rate per cycle, and intersect the break-even line at smaller \(p\). This shows that stronger nonlocal noise lowers the pseudo-threshold. This trend is consistent with the ansatz: \(\alpha\) changes the coefficient functions \(c_0(\alpha)\), \(c_1(\alpha)\), and \(c_2(\alpha)\), while the leading BB exponent remains fixed.
\vspace{2pt}
Table~\ref{tab:quadratic_alpha_coefficients_values} reports the evaluated values of the coefficient functions $c_0(\alpha)$, $c_1(\alpha)$, and $c_2(\alpha)$ at the simulated values of $\alpha$, thereby specifying the fitted curve for each nonlocal noise setting. Table~\ref{tab:quadratic_alpha_threshold_summary} summarizes the resulting pseudo thresholds together with the logarithmic goodness of fit metrics ($RMSE$ and $R^{2}$). RMSE measures the typical absolute fitting error, while \(R^2\) measures the variance explained by the fit. The corresponding \(\mathrm{log\,RMSE}\) and \(\mathrm{log}\,R^2\) are computed after taking logarithms of the logical error rates. Smaller \(\mathrm{log\,RMSE}\) and \(\mathrm{log}\,R^2\) closer to unity indicate a more accurate fit~\cite{montgomery2021introduction}.

\vspace{2pt}

The extracted values show a clear monotonic decrease in $p_0(\alpha)$, from $0.006343$ at $\alpha=1$ to $0.001651$ at $\alpha=7$, which quantitatively confirms the increasing penalty associated with larger nonlocal error strength. The reported logarithmic fit metrics indicate that the quadratic $\alpha$ dependent ansatz provides a consistently accurate description of the simulated data over the fitting region.

\begin{table}[h!]
    \centering
    \caption{Evaluated values of the coefficient functions $c_0(\alpha)$, $c_1(\alpha)$, and $c_2(\alpha)$ at the simulated values of $\alpha$ for 12 QPUs.}
    \label{tab:quadratic_alpha_coefficients_values}
    \begin{tabular}{c c c c}
        \hline
        $\alpha$ & $c_0(\alpha)$ & $c_1(\alpha)$ & $c_2(\alpha)$ \\
        \hline
        1 & 17.532268 & 1541.451577 & $-1.139137 \times 10^{5}$ \\
        3 & 19.827294 & 2734.133122 & $-3.199404 \times 10^{5}$ \\
        5 & 21.452477 & 4109.170765 & $-7.285799 \times 10^{5}$ \\
        7 & 22.407817 & 5666.564507 & $-1.339832 \times 10^{6}$ \\
        \hline
    \end{tabular}
\end{table}

\begin{table}[h!]
    \centering
    \caption{Summary of the $\alpha$ dependent pseudo-threshold $p_0(\alpha)$ and logarithmic goodness-of-fit metrics for 12 QPUs.}
    \label{tab:quadratic_alpha_threshold_summary}
    \begin{tabular}{c c c c}
        \hline
        $\alpha$ & $p_0(\alpha)$ & $\log \mathrm{RMSE}$ & $\log R^2$ \\
        \hline
        1 & 0.006343 & 0.036220 & 0.999859 \\
        3 & 0.003286 & 0.304231 & 0.989884 \\
        5 & 0.002198 & 0.086283 & 0.998407 \\
        7 & 0.001651 & 0.045167 & 0.999109 \\
        \hline
    \end{tabular}
\end{table}

The logarithmic fit metrics in Table~\ref{tab:quadratic_alpha_threshold_summary} quantify how well the quadratic $\alpha$ dependent ansatz captures the simulated data within the fitted region. Small $\log \mathrm{(RMSE)}$ and $\log (R^2)$ values close to unity indicate that the ansatz provides an accurate and consistent description of the observed logical error scaling.


\subsection{Quadratic $\alpha$ dependent pseudo threshold fits for 6 QPUs}

\begin{figure}[htbp]
    \centering
    \includegraphics[width=0.5\textwidth]{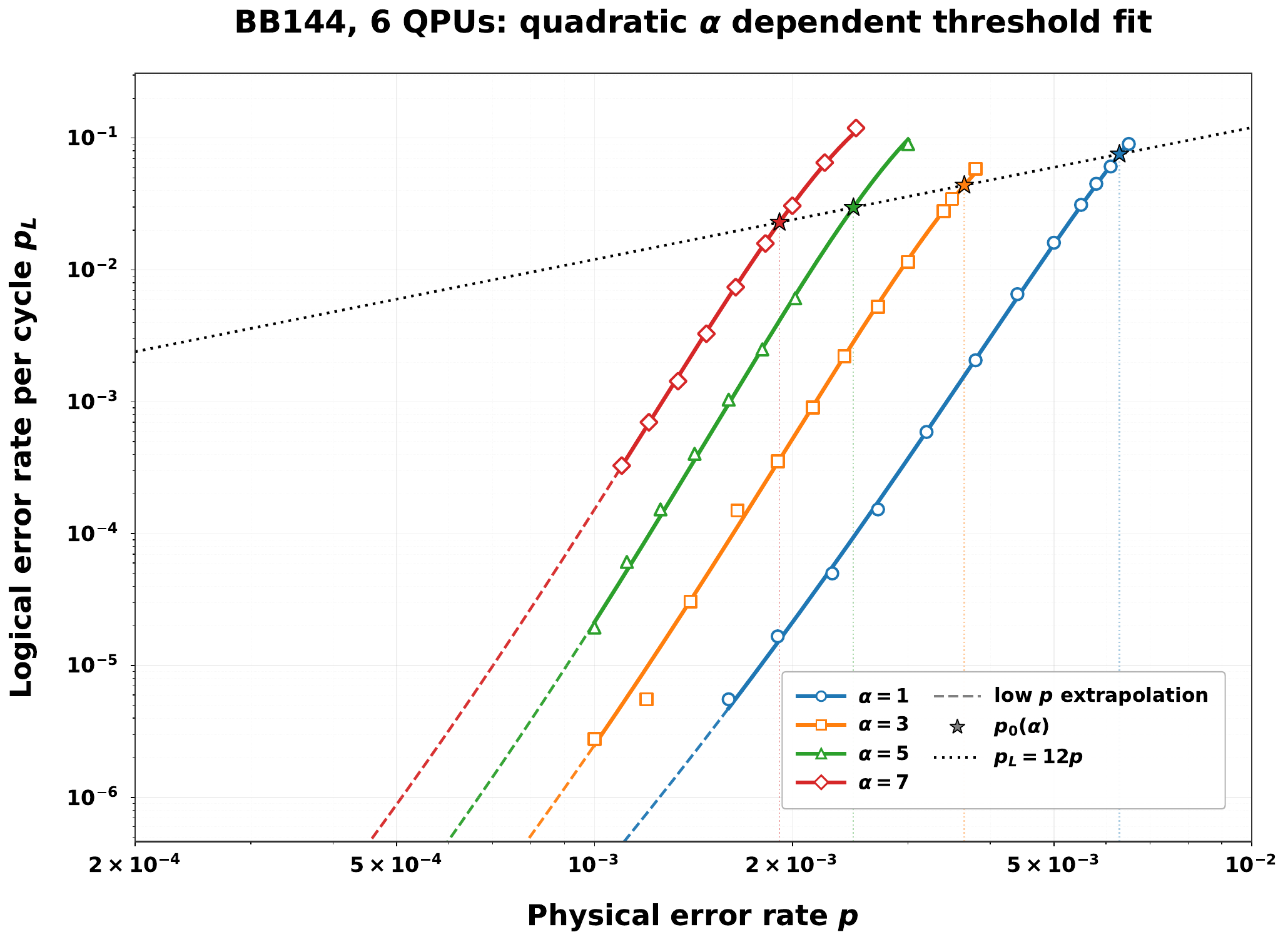}
    \caption{BB144 threshold fitting with quadratic $\alpha$ dependence for 6 QPUs. The logical error rate per cycle $p_L$ is plotted against the physical error rate $p$ for different values of $\alpha$. The solid colored curves show the fitted ansatz in the fitting region, while the dashed colored segments indicate extrapolation to lower $p$. The dotted black line denotes the break even condition $p_L = 12p$, and the asterisks mark the corresponding pseudo thresholds $p_0(\alpha)$.}
    \label{fig:bb144_qpu6_threshold_fit}
\end{figure}

Figure~\ref{fig:bb144_qpu6_threshold_fit} shows the corresponding quadratic \(\alpha\)-dependent fits for the BB144 code distributed across 6 QPUs. As in the 12-QPU case, the fits show similar behavior: increasing \(\alpha\) shifts the fitted curves upward and moves their intersections with the break-even line \(p_L=12p\) to smaller \(p\), indicating a reduction in the pseudo-threshold. For 6 QPUs, the extracted pseudo-threshold decreases monotonically from \(0.006291\) at \(\alpha=1\) to \(0.001912\) at \(\alpha=7\).

\vspace{2pt}

Table~\ref{tab:quadratic_alpha_coefficients_values_qpu6} lists the evaluated coefficient functions $c_0(\alpha)$, $c_1(\alpha)$, and $c_2(\alpha)$ for the fitted ansatz, while Table~\ref{tab:quadratic_alpha_threshold_summary_qpu6} summarizes the resulting values of $p_0(\alpha)$ together with the logarithmic goodness of fit metrics. 


\begin{table}[h!]
    \centering
    \caption{Evaluated values of the coefficient functions $c_0(\alpha)$, $c_1(\alpha)$, and $c_2(\alpha)$ at the simulated values of $\alpha$ for 6 QPUs.}
    \label{tab:quadratic_alpha_coefficients_values_qpu6}
    \begin{tabular}{c c c c}
        \hline
        $\alpha$ & $c_0(\alpha)$ & $c_1(\alpha)$ & $c_2(\alpha)$ \\
        \hline
        1 & 18.162811 & 1235.887120 & $-8.030391 \times 10^{4}$ \\
        3 & 18.998369 & 3025.031353 & $-3.833232 \times 10^{5}$ \\
        5 & 20.236679 & 4242.541588 & $-6.939576 \times 10^{5}$ \\
        7 & 21.877743 & 4888.417825 & $-1.012207 \times 10^{6}$ \\
        \hline
    \end{tabular}
\end{table}

\begin{table}[h!]
    \centering
    \caption{Summary of the $\alpha$ dependent pseudo threshold $p_0(\alpha)$ and logarithmic goodness of fit metrics for 6 QPUs.}
    \label{tab:quadratic_alpha_threshold_summary_qpu6}
    \begin{tabular}{c c c c}
        \hline
        $\alpha$ & $p_0(\alpha)$ & $\log \mathrm{RMSE}$ & $\log R^2$ \\
        \hline
        1 & 0.006291 & 0.075884 & 0.999461 \\
        3 & 0.003653 & 0.183672 & 0.996831 \\
        5 & 0.002477 & 0.095032 & 0.998594 \\
        7 & 0.001912 & 0.038384 & 0.999605 \\
        \hline
    \end{tabular}
\end{table}

\subsection{Quadratic $\alpha$ dependent pseudo threshold fits for 4 QPUs}



\begin{figure}[htbp]
    \centering
    \includegraphics[width=0.5\textwidth]{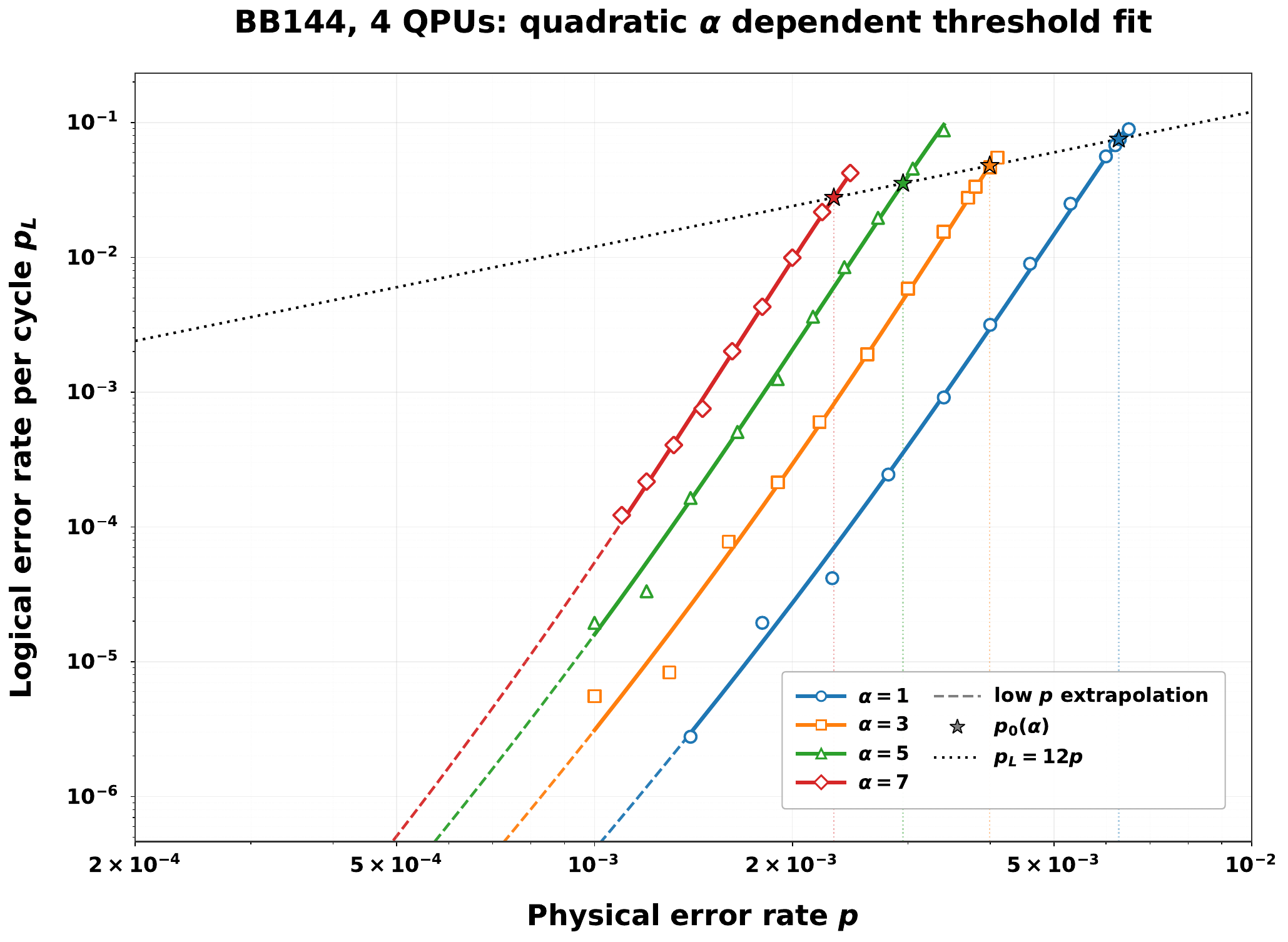}
    \caption{BB144 threshold fitting with quadratic $\alpha$ dependence for 4 QPUs. The logical error rate per cycle $p_L$ is plotted against the physical error rate $p$ for different values of $\alpha$. The solid colored curves show the fitted ansatz in the fitting region, while the dashed colored segments indicate extrapolation to lower $p$. The dotted black line denotes the break even condition $p_L = 12p$, and the asterisks mark the corresponding pseudo thresholds $p_0(\alpha)$.}
    \label{fig:bb144_qpu4_threshold_fit}
\end{figure}

Figure~\ref{fig:bb144_qpu4_threshold_fit} shows the quadratic \(\alpha\) dependent fits for the BB144 code distributed across 4 QPUs. As \(\alpha\) increases, the curves shift upward and cross \(p_L=12p\) at smaller \(p\), reducing the pseudo threshold from \(0.006276\) to \(0.002313\).

\vspace{2pt}

Table~\ref{tab:quadratic_alpha_coefficients_values_qpu4} lists the evaluated coefficient functions $c_0(\alpha)$, $c_1(\alpha)$, and $c_2(\alpha)$ for the fitted ansatz, while Table~\ref{tab:quadratic_alpha_threshold_summary_qpu4} summarizes the resulting values of $p_0(\alpha)$ together with the logarithmic goodness of fit metrics. 


\begin{table}[h!]
    \centering
    \caption{Evaluated values of the coefficient functions $c_0(\alpha)$, $c_1(\alpha)$, and $c_2(\alpha)$ at the simulated values of $\alpha$ for 4 QPUs.}
    \label{tab:quadratic_alpha_coefficients_values_qpu4}
    \begin{tabular}{c c c c}
        \hline
        $\alpha$ & $c_0(\alpha)$ & $c_1(\alpha)$ & $c_2(\alpha)$ \\
        \hline
        1 & 18.987403 & 874.153832 & $-4.327385 \times 10^{4}$ \\
        3 & 20.611695 & 1330.226302 & $-8.441741 \times 10^{4}$ \\
        5 & 21.639868 & 2078.467291 & $-2.272227 \times 10^{5}$ \\
        7 & 22.071923 & 3118.876802 & $-4.716897 \times 10^{5}$ \\
        \hline
    \end{tabular}
\end{table}

\begin{table}[h!]
    \centering
    \caption{Summary of the $\alpha$ dependent pseudo threshold $p_0(\alpha)$ and logarithmic goodness of fit metrics for 4 QPUs.}
    \label{tab:quadratic_alpha_threshold_summary_qpu4}
    \begin{tabular}{c c c c}
        \hline
        $\alpha$ & $p_0(\alpha)$ & $\log \mathrm{RMSE}$ & $\log R^2$ \\
        \hline
        1 & 0.006276 & 0.177759 & 0.997379 \\
        3 & 0.003993 & 0.265740 & 0.993052 \\
        5 & 0.002948 & 0.176814 & 0.995989 \\
        7 & 0.002313 & 0.075022 & 0.998526 \\
        \hline
    \end{tabular}
\end{table}

\subsection{Comparison across QPU partitions}

\begin{figure}[htbp]
    \centering
    \includegraphics[width=0.45\textwidth]{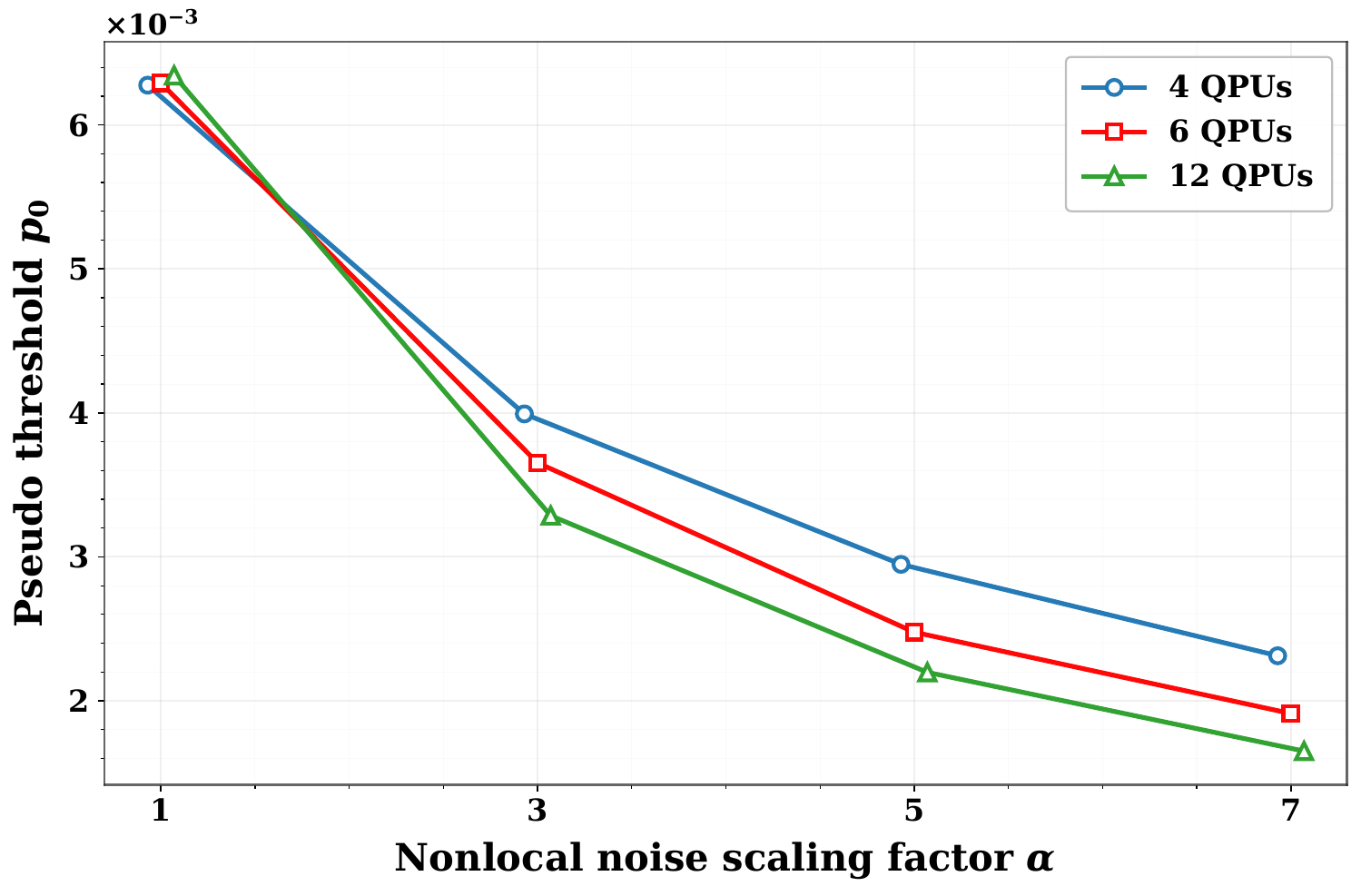}
    \caption{Comparison of the extracted pseudo threshold $p_0(\alpha)$ as a function of the nonlocal noise scaling factor $\alpha$ for the 4, 6, and 12 QPU partitions of the BB144 code.}
    \label{fig:bb144_pseudothreshold_vs_alpha}
\end{figure}

Figure~\ref{fig:bb144_pseudothreshold_vs_alpha} compares the extracted pseudo-threshold $p_0(\alpha)$ across the 4, 6, and 12 QPU partitions of the BB144 code. For all three partitions, the pseudo threshold decreases monotonically as the nonlocal noise scaling factor $\alpha$ increases. At $\alpha=1$, the pseudo threshold values are very similar across the three partitions. For \(\alpha=1\), Bravyi \emph{et al.} report \(p_0\simeq 0.0065\), while our fitted values for BB144 are \(0.006343\), \(0.006276\), and \(0.006291\), showing close agreement; the small differences are expected from the use of different sampled \(p\) points, and different numbers of trials~\cite{Bravyi2024}.

\vspace{2pt}


For $\alpha \ge 3$, the 4-QPU partition consistently yields the highest pseudo-threshold, followed by the 6-QPU partition, with the 12-QPU partition exhibiting the lowest performance. This trend confirms that an increased number of non-local gates adversely impacts the pseudo-threshold, primarily because the 12-QPU partition necessitates more frequent inter-QPU communication. While all partitions demonstrate degradation as $\alpha$ increases, minimizing partitioning is more effective at preserving the pseudo-threshold by reducing the number of non-local gates. These findings underscore the inherent \textit{scalability-threshold} trade-off in distributed quantum computing, suggesting that modularity entails a cost in threshold performance and that future advancements in quantum interconnects will be essential to mitigate these effects.

\section{Conclusion and Future Works}
\label{sec:conclusion}

In this work, we studied distributed realizations of the $[\![144,12,12]\!]$ Bivariate bicycle code across 4, 6, and 12 QPUs which can be realized in a modular star network architecture. We considered a circuit-level noise model that distinguishes local and nonlocal CNOT gates through a nonlocal noise scaling factor $\alpha$, and used Monte Carlo simulations together with BP+OSD decoding to characterize the resulting logical error rates and pseudo-threshold behavior. To analyze these results, we also introduced a quadratic $\alpha$-dependent extension of the BB Code ansatz, which enabled extrapolation of low physical error rates.

Our results show that the pseudo-threshold decreases monotonically with increasing $\alpha$ for all considered QPU partitions, confirming the growing penalty associated with stronger nonlocal noise. We also found that, coarser partitions become more favorable as $\alpha$ increases, with the 4 QPU case  outperforming the 6 and 12 QPU cases for $\alpha \ge 3$. Our findings provide insight into the tradeoff between modular partitioning and nonlocal error cost in distributed implementations of a BB code.

Future work can examine the possibility of Bell pair mediated operations in BB codes introducing correlated errors and distance reducing hook errors, where a single circuit error spreads to multiple data qubits. It would also be useful to study how modified syndrome extraction procedures could mitigate these effects. Other important directions include studying severe error mechanisms such as catastrophic qubit failures and leakage. Finally, strategies for distributed realizations of qLDPC codes could optimize qubit placement and entangled pair scheduling while accounting for hardware constraints, such as limited connectivity within QPUs, sparse network connectivity between QPUs, and rate constrained entanglement links.

\section*{Acknowledgment}

KPS thanks the U.S. Department of Energy, Office of Science, Advanced Scientific Computing Research (ASCR) program, for support under Award Number DE-SC0026264. KPS also gratefully acknowledges Cisco Gift Award in support of distributed quantum computing research. This research was supported in part by the University of Pittsburgh Center for Research Computing and Data (CRCD), \texttt{RRID:SCR\_022735}, through the resources provided. Specifically, this work used the H2P cluster, which is supported by NSF award number \texttt{OAC-2117681}. NKC thanks the CRCD team.

\bibliographystyle{IEEEtran}
\bibliography{references}

\end{document}